\documentclass[10pt]{article}
\usepackage{amsmath}
\usepackage{amsfonts}
\usepackage{epsfig}
\usepackage{latexsym}
\usepackage{url}

\def\noi{\noindent}

\def\ep{\varepsilon}
\def\la{\lambda}
\def\tr{\hbox{tr}}

\def\bra{\langle}
\def\ket{\rangle}
\def\str{\rightarrow}

\def\pmx{\begin{pmatrix}}
\def\emx{\end{pmatrix}}

\def\R{\mathbb{R}}
\def\C{\mathbb{C}}
\def\dg{\dagger}

\date{}

\begin{document}

\title {Still more on norms of completely positive maps}
\author{Stanislaw J. Szarek (Cleveland and Paris)}

\maketitle

%\begin{abstract}
%\end{abstract}
Let $\mathcal{M}_n$ denote the space of $n \times n$ (real or complex) 
matrices and, for $A \in \mathcal{M}_n$ and $p\geq 1$,  let
$\|A\|_p:= (\tr(A^\dg A)^{p/2})^{1/p}$ be the Schatten $p$-norm of $A$, 
with the limit case $p=\infty$ corresponding to the usual operator norm.
Further, if  $\Phi: \mathcal{M}_m \str \mathcal{M}_n$ is a 
linear map and $p, q \in [1, \infty]$, we consider
\begin{equation} \label{defpq}
\|\Phi\|_{p \str q} := \max \{ \|\Phi(\sigma)\|_q \ : \ 
\sigma \in \mathcal{M}_m, \|\sigma\|_p \leq 1\} , 
\end{equation}
 i.e., the norm of
$\Phi$ as an operator between the %corresponding two 
normed spaces
$(\mathcal{M}_m,\|\cdot\|_p)$ and $(\mathcal{M}_n,\|\cdot\|_q)$.
%We want to compare $\|\Phi\|_{\al \str \be}$ 
Such quantities were studied (in the context of 
quantum information theory) in \cite{kr}, where the 
question was raised under what conditions 
 (\ref{defpq}) coincides   with the {\em a priori} smaller norm 
\begin{equation} \label{defpqh}
\|\Phi\|_{p \str q}^H := \max \{ \|\Phi(\sigma)\|_q \ : \ 
\sigma \in \mathcal{M}_m, \sigma = \sigma^\dg, \|\sigma\|_p \leq 1\}
\end{equation}
of the restriction of $\Phi$ to the (real
linear) subspace of {\em Hermitian} matrices and, in particular, 
whether this holds if $\Phi$ is {\em completely positive}.
[Note that if  $\Phi$ is just {\em positivity preserving}, it maps Hermitian
matrices to Hermitian matrices.]
The latter was subsequently confirmed in \cite{watrous, aud}, the first of which
also contains an assortment of examples showing when such equalities may or may
not hold (see also \cite{djkr}). Here we provide one more
proof. More precisely, we will show 

\medskip \noi {\bf Proposition }  {\em If $\Phi$ is $2$-positive, then }
$\|\Phi\|_{p \str q} = \|\Phi\|_{p \str q}^H$. {\sl Moreover, 
similar equality holds if the domain and the range of $\Phi$ are endowed with
any unitarily invariant norms.}

\medskip \noi 
Recall that a norm $\|\cdot\|$ on $\mathcal{M}_n$ is called 
{\em unitarily invariant} if $\|UAV\| = \|A\|$ for any $A \in
\mathcal{M}_n$  and any $U, V \in U(n)$ (resp., $O(n)$ in the real case); 
see \cite{s,hj}.  This is
equivalent to requiring that the norm of a matrix depends only on its {\em
singular values} (called in some circles ``Schmidt coefficients"). 
Besides using a slightly weaker hypothesis and yielding a slightly more general
assertion,
%(at least some of these features can undoubtedly be obtained from the
%earlier proofs; e.g., an analysis of \cite{aud} 
%and its references shows that 
%in fact only 2-positivity is needed there),
the argument we present is self-contained and uses only
definitions and elementary facts and concepts from linear algebra, of which
the most sophisticated is the singular value decomposition. It may thus be
argued that it is the ``right" proof. 
[Note that an analysis of \cite{aud} and its references shows that 
in fact only 2-positivity is needed there, too.]

\medskip \noi  {\em Proof }
For clarity, we will consider first the case when $p=1$, i.e., when the domain of
$\Phi$ is endowed with the trace class norm. 
In this case the extreme points of the respective unit balls 
(on which the maxima 
in (\ref{defpq}) and (\ref{defpqh}) are
necessarily achieved) are particularly simple:  they are rank one operators.
Accordingly, the question reduces to showing that 
\begin{equation} \label{eq1}
\max_{|u| = |v| = 1} \|\Phi(|v\ket \bra u |) \|_q\ 
\leq \ \max_{|u| = 1} \|\Phi(|u\ket \bra u |)\|_q ,
\end{equation}
where $u, v \in \C^m$ (or $\R^m$, depending on the context) and 
$|\cdot |$ is the Euclidean norm. Given such $u, v$, consider the block  matrix 
$M_{u,v}=\begin{bmatrix}|u\ket \bra u | & |u \ket \bra v |\\ |v\ket \bra u | &
|v\ket\bra v |\end{bmatrix} \in \mathcal{M}_{2m}$ and
note that  $M_{u,v} = | \xi\ket \bra \xi |$ where
$| \xi\ket = (|u \ket ,|v\ket ) \in \C^r \oplus \C^r$
(in particular $M_{u,v} \geq 0$).  
Considering $M_{u,v}$ as an element of 
$\mathcal{M}_m \otimes \mathcal{M}_2$ and appealing to $2$-positivity of  $\Phi$ we
deduce that
$(\Phi \otimes Id_{\mathcal{M}_2})(M_{u,v}) = 
\begin{bmatrix}\Phi(|u\ket \bra u |) & \Phi(|u \ket \bra v |)\\ 
\Phi(|v\ket \bra u |) & \Phi(|v\ket \bra v |)\end{bmatrix}~
\geq~0$. The conclusion now follows from the following  
lemma (see, e.g., \cite{hj}, Theorem 3.5.15; for completeness we include
a proof at the end of this note).

\medskip \noi {\bf Lemma } {\sl Let $A, B, C \in \mathcal{M}_r$ be such that 
the $2r \times 2r$ block matrix 
$M = \begin{bmatrix}A & B\\ B^\dg & C\end{bmatrix}$ is positive
semi-definite, and let  $\|\cdot\|$ be a unitarily invariant norm 
 on} $\mathcal{M}_r$. {\em Then }
$%$
\|B\|^2 \leq \|A\| \, \|C\| \, .
$%$

\bigskip \noi The case of arbitrary $p \in [1,\infty]$ is almost as
simple.   First, for $\sigma \in \mathcal{M}_m$ with $\|\sigma\|_p \leq 1$ we
consider  the positive semi-definite matrix $M_\sigma =\begin{bmatrix}(\sigma
\sigma^\dg)^{1/2} & \sigma\\ 
\sigma^\dg & (\sigma^\dg \sigma)^{1/2} \end{bmatrix}$. 
[Positivity is seen, e.g., by writing down the singular value
decompositions of the entries and expressing $M_\sigma$ as a positive
linear combination  of matrices of the type $M_{u,v}$ considered above.]
Since unitarily invariant norms depend only on singular values of a
matrix, we have
$\|(\sigma \sigma^\dg)^{1/2}\|_p= \| (\sigma^\dg \sigma)^{1/2}\|_p 
= \|\sigma\|_p \leq 1$. 
On the
other hand, arguing as in the special case $p=1$, we deduce 
from the Lemma that $\|\Phi(\sigma)\|_q^2 \leq \|\Phi((\sigma
\sigma^\dg)^{1/2})\|_q\, \|\Phi((\sigma^\dg \sigma)^{1/2})\|_q
\leq \big(\|\Phi\|_{p \str q}^H\big)^2$, and the conclusion follows by
taking the maximum over $\sigma$.
The proof for general unitarily invariant norms %on the domain of $\Phi$
is the same. 
\hfill$\square$

\medskip \noi {\em Proof of the Lemma } 
[Written for $\|\cdot\| = \|\cdot\|_q$, but the general
case works in the same way.] 
Let  
$B = \sum_{j=1}^r \la _j |\varphi_j\ket \bra \psi_j|$ be the Schmidt 
decomposition. Consider the orthonormal basis of $\C ^{2r}$ which is a 
concatenation
of $(| \varphi_j \ket)$ and $(| \psi_j \ket)$. The representation of $M$ in
that  basis is 
$$
M' := \begin{bmatrix} \left(\bra \varphi_j |A| \varphi_k \ket\right)_{j,k=1}^r &
{\rm Diag}(\la )\\  
{\rm Diag}(\la ) & \left(\bra \psi_j |C| \psi_k
\ket\right)_{j,k=1}^r\end{bmatrix},$$ where ${\rm Diag}(\mu)$ is the
diagonal  matrix with the sequence $\mu = (\mu_j)$ on the diagonal.
Given $j \in \{1,\ldots,r\}$, the $2 \times 2$ matrix 
$\begin{bmatrix}\bra \varphi_j |A| \varphi_j \ket & \la _j\\  
\la _j & \bra \psi_j |C|\psi_j\ket\end{bmatrix}$ is a minor of $M'$ and
hence positive semi-definite,
and so 
$\la _j \leq 
\sqrt{\bra \varphi_j |A| \varphi_j \ket \bra \psi_j |C|\psi_j\ket}
\leq (\bra \varphi_j |A| \varphi_j \ket + \bra \psi_j |C|\psi_j\ket)/2$ . 
Consequently
\begin{eqnarray} \label{BAC}
\|B\|_q= \big(\sum_j \la_j^q \big)^{1/q} & \leq &
\big(\big(\sum_j \bra \varphi_j |A| \varphi_j \ket^q \big)^{1/q} +
\big(\sum_j
\bra \psi_j |C|\psi_j\ket^q \big)^{1/q}\big)/2 \nonumber \\
&\leq &(\|A\|_q +\|C\|_q)/2 .
\end{eqnarray}
The last inequality in (\ref{BAC}) follows from the well-known fact that,
for {\em any } square matrix 
$S=(S_{jk})$, $\|S\|_q \geq \big(\sum_j S_{jj}^q\big)^{1/q}$  
 (which in turn is a consequence of $(S_{jk} \delta_{jk})$, the
diagonal part of $S$, being the average of
${\rm Diag}(\ep) \, S \, {\rm Diag}(\ep)$, where 
$\ep=(\ep_{j})$ varies over all choices of $\ep _j =\pm 1$). 
The bound from (\ref{BAC}) is already sufficient to
prove (\ref{eq1}) (and the Proposition). To obtain the stronger statement
from the Lemma we use the inequality $ab \leq \frac12 (ta+b/t)$  (for
$t>0$, instead of 
$ab \leq \frac12 (a+b)$) to obtain $\|B\|_q \leq \frac 12 (t\|A\|_q
+\|C\|_q/t)$, and then specify the optimal value $t=(\|C\|_q/\|A\|_q)^{1/2}$.
Passing to a generic unitarily invariant norm requires just replacing 
everywhere $\big(\sum_j \mu_j^q \big)^{1/q}$ by $\|{\rm Diag}(\mu)\|$; 
equalities such as $\|B\|=\|{\rm Diag}(\la )\|$ or 
$\|A\|=\|\left(\bra \varphi_j |A| \varphi_k \ket\right)_{j,k=1}^r\|$
just express the unitary invariance of the norm.
\hfill$\square$

\bigskip 
\noindent The author thanks K. Audenaert and M. B. Ruskai for comments on
the first version of this note.

\footnotesize{
Universit\'e Pierre et Marie Curie-Paris 6, 
UMR 7586-Institut de Math\'ematiques, 
Analyse Fonctionnelle, BC 186, 75252 Paris, France  \ {\sl and}\\
Case Western Reserve University, 
Department of Mathematics, Cleveland, Ohio 44106-7058, U.S.A.\\
Email: szarek@math.jussieu.fr
}

\end{document}